\documentclass[pra,letterpaper,aps,10pt,superscriptaddress,twocolumn,floatfix,showpacs]{revtex4-1}
\usepackage{amsmath,graphicx,amssymb,braket,xcolor,subfigure,upgreek}
\usepackage[colorlinks, linkcolor=blue, citecolor=blue, urlcolor=blue, breaklinks=true]{hyperref}

\newcommand{\abs}[1]{\left\lvert#1\right\rvert}
\newcommand{\eref}[1]{Eq.~(\ref{#1})}

\newcommand{\fref}[1]{Fig.~\ref{#1}}

\begin{document}

\title{Enhanced optomechanical readout using optical coalescence}

\author{Claudiu Genes}
\email[Email address:\ ]{claudiu.genes@uibk.ac.at}
\affiliation{ISIS (UMR 7006) and IPCMS (UMR 7504), Universit\'{e} de Strasbourg and CNRS, Strasbourg, France}
\affiliation{Institut f\"ur Theoretische Physik, Universit\"at Innsbruck, Technikerstrasse 25, A-6020 Innsbruck, Austria}
\author{Andr\'{e} Xuereb}
\affiliation{Centre for Theoretical Atomic, Molecular and Optical Physics, School of Mathematics and Physics, Queen's University
Belfast, Belfast BT7\,1NN, United Kingdom}
\affiliation{Department of Physics, University of Malta, Msida MSD\,2080, Malta}
\author{Guido Pupillo}
\affiliation{ISIS (UMR 7006) and IPCMS (UMR 7504), Universit\'{e} de Strasbourg and CNRS, Strasbourg, France}
\author{Aur\'{e}lien Dantan}
\affiliation{QUANTOP, Danish National Research Foundation Center for Quantum Optics, Department of Physics and Astronomy, University of Aarhus, 8000 Aarhus C, Denmark}

\date{\today}

\begin{abstract}
We present a scheme to strongly enhance the readout sensitivity of the squared displacement of a mobile scatterer placed in a Fabry--P\'erot cavity. We investigate the largely unexplored regime of cavity electrodynamics in which a highly-reflective element positioned between the end-mirrors of a symmetric Fabry--P\'erot resonator strongly modifies the cavity response function, such that two longitudinal modes with different spatial parity are brought close to frequency degeneracy and interfere in the cavity output field. In the case of a movable middle reflector we show that the interference in this generic `optical coalescence' phenomenon gives rise to an enhanced frequency shift of the peaks of the cavity transmission that can be exploited in optomechanics.
\end{abstract}


\maketitle

A Fabry--P\'erot resonator containing a scattering element between its two end-mirrors represents a paradigmatic system for fundamental light--matter interaction studies, as investigated, e.g., in cavity quantum electrodynamics (CQED)~\cite{Berman1994,*Haroche2006,*Kimble2008} and cavity optomechanics~\cite{Aspelmeyer2013}. The theoretical description of light in high-finesse cavities, i.e., cavities that possess resonances spaced significantly further apart than their individual widths, is well-established and the modification of the bare cavity response induced by the presence of a scatterer has been exploited in different regimes, distinguished by the scatterer's reflectivity. In a first regime, a low-reflectivity scatterer, e.g., an atom, is coupled to a single optical mode that spans the entire cavity. While this atom can be strongly coupled to the mode, its presence does not modify the field mode function appreciably~\cite{Xuereb2012d}. At the opposite end of the scale, massive mobile scatterers can substantially alter the cavity field modes; as studied in optomechanics this, allows for a rich variety of tools to measure and control mechanical motion~\cite{Aspelmeyer2013}. A noteworthy example is that of a harmonically-bound central element, such as a thin, partially transmitting membrane~\cite{Thompson2008}, placed in between highly reflecting mirrors, for which various forms of optomechanical coupling can be engineered~\cite{Thompson2008,Jayich2008,Sankey2010}. Such a system has been used to experimentally observe backaction optomechanical cooling~\cite{Thompson2008,Wilson2009,Kemiktarak2012,*Kemiktarak2012b,Karuza2013}, radiation pressure shot-noise~\cite{Purdy2013}, ponderomotive squeezing~\cite{Purdy2013b}, and optomechanically-induced transparency~\cite{Karuza2013b}, and proposals exist for, e.g., the observation of jumps in the occupation number of the oscillator~\cite{Thompson2008,Clerk2010,Ludwig2012}, tests of the Landau--Zener effect~\cite{Heinrich2010,*Miladinovic2011}, the generation of nonclassical states of motion~\cite{Bose1997,*Kleckner2008,*Hartmann2008,*Paternostro2011,*Buchmann2012}, quantum information processing~\cite{Stannigel2010,Stannigel2012}, or the coupling to cold atoms~\cite{Hammerer2009,Camerer2011,Genes2011,Vogell2013}.

\begin{figure}[t]
 \includegraphics[width=0.99\columnwidth]{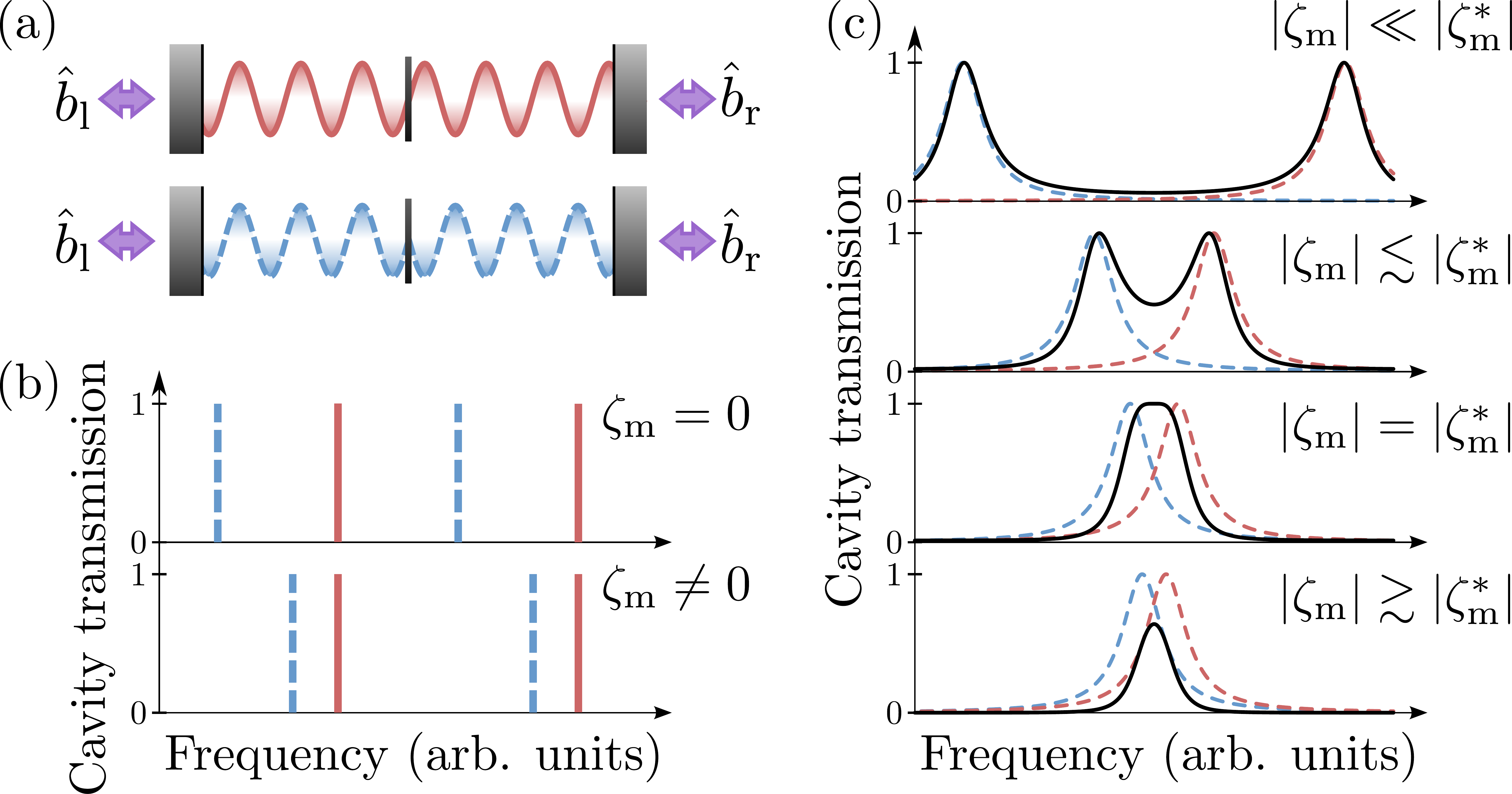}
 \caption{(Color online) (a)~Optical cavity with a middle reflector of polarizability $\zeta_\mathrm{m}$, accommodating odd (red) and even (blue) modes that couple to outside ``left'' and ``right'' modes. (b)~Cavity transmission spectra over three free spectral ranges, illustrating the strong shift of the odd resonances toward the even ones as the reflectivity is increased. (c)~Transmission spectra (solid black curves) for increasing $\abs{\zeta_\mathrm{m}}$. The dashed curves denote the ``bare'' resonances. `Mode-pulling' effects can be seen by comparing the positions of these resonances and the peaks of the transmission. At coalescence ($\zeta_\mathrm{m}^\ast$ is defined in the text) one finds a single peak with close to unity transmission over an increased bandwidth. Past this point the maximum transmission is reduced below unity.}
 \label{fig:CavityModes}
\end{figure}

This generic membrane-in-the-middle system is generally described by means of two spatially-separated cavity modes coupled through photon tunneling. In the optomechanical situations commonly considered, the reflectivity of the middle scatterer is typically much lower than that of the end mirrors. In this work we use the transfer matrix formalism, as employed recently to unify these two regimes~\cite{Xuereb2012d}, to go beyond this case and address the situation where the reflectivity of the middle mirror is increased well past that of the cavity mirrors. This situation could be relevant for, e.g., highly-reflective membranes~\cite{Kemiktarak2012,*Kemiktarak2012b} or arrays of membranes~\cite{Xuereb2012c}. We focus particularly on the situation where adjacent cavity resonances are brought very close together. Under such circumstances, we find that the transmission resonances in the \emph{output} field are strongly pulled together, leading eventually to an \emph{optical coalescence} phenomenon, where pairs of adjacent resonances with different spatial parity become almost degenerate. The interference between these modes in the output fields modifies the cavity response in a non-trivial fashion. In particular, close to the coalescence point, high transmission can be achieved over a bandwidth larger than the bare cavity linewidth. While this mechanism alone could be relevant for practical applications of optical cavities in interferometry and, e.g., the realization of white-light cavities~\cite{Wicht1997,*Pati2007,*Savchenkov2006}, we show that it also has interesting implications for cavity optomechanics. Owing to this coalescence-induced pulling of the transmission resonance frequencies, we show that the intrinsic quadratic optomechanical coupling translates to a strongly enhanced optomechanical readout sensitivity as compared to the situations envisaged in standard optomechanical models~\cite{Aspelmeyer2013,Thompson2008,Sankey2010,Xuereb2012d}.
\par
\noindent\emph{Model.}---Let us consider a cavity of length $L$ consisting of two mirrors with polarizability $\zeta<0$, related to the amplitude reflectivity through {$r=-{i\zeta}/{(1-i\zeta)}$}, and a middle reflector with polarizability $\zeta_\mathrm{m}$, cf.\ \fref{fig:CavityModes}(a). We assume that the mirrors are lossless (i.e., described by real polarizabilities) and that the middle reflector has a thickness smaller than the wavelengths considered, themselves much smaller than $L$. We make use of the transfer matrix formalism for one-dimensional scatterers to solve the Helmholtz equation~\cite{Deutsch1995,Xuereb2012d}, and compute the transmission function of the system for general values of $\zeta$ and $\zeta_\mathrm{m}$. For an empty cavity with sufficiently good mirrors the transmission spectrum consists of well-separated Lorentzian peaks with frequencies given by $\omega_{n}=(c/L)\bigl[(n-1)\pi+\cos^{-1}\bigl(\zeta/\sqrt{1+\zeta^2}\bigr)\bigr]$ ($n=1,2,...$) and linewidth $\kappa=c/(2L\abs{\zeta}\sqrt{1+\zeta^2})$. In the presence of the middle reflector, positioned exactly at the center of the cavity, adjacent resonance frequencies are shifted closer together by an amount that depends on its position and polarizability [\fref{fig:CavityModes}(b)]. When the reflector lies exactly at the center of the cavity, as shown in \fref{fig:CavityModes}(a), the peaks quantified by an even mode number (`e', frequency $\omega_\mathrm{e}$) are shifted strongly towards the odd-numbered (`o', $\omega_\mathrm{o}$) peaks:
\begin{equation*}
\omega_{\mathrm{o}}-\omega_{\mathrm{e}}\approx\tfrac{c}{L}\tan^{-1}\bigl[2\zeta_\mathrm{m}/(\zeta_\mathrm{m}^2-1)\bigr]\equiv2\delta\,.
\end{equation*}
For modest middle-mirror reflectivities the transmission peaks have non-overlapping Lorentzian profiles, which ensures that they indeed occur at the cavity resonances to a good approximation [see \fref{fig:CavityModes}(c)]. However, as $\abs{\zeta_{m}}$ is increased, the frequency separation between $\omega_{\mathrm{o}}$ and $\omega_{\mathrm{e}}$ can become comparable to their linewidth and the two modes begin to interfere with each other in the output field. In this regime the peaks of the transmission are no longer located at the cavity resonances but at $\tilde{\omega}_{\mathrm{e}}=\tfrac{c}{L}\bigl[(2n)\pi-\epsilon_-\bigr]$ and $\tilde{\omega}_{\mathrm{o}}=\tfrac{c}{L}\bigl[(2n+1)\pi-\epsilon_+\bigr]$ where
\begin{equation*}
\cos\epsilon_{\pm }=\tfrac{\zeta_\mathrm{m}(2\zeta^2+1)(\zeta\zeta_{m}-1)\pm(\zeta+\zeta_{m})\sqrt{4\zeta^2(\zeta^2+1)-\zeta_\mathrm{m}^2}}{2\zeta(\zeta^2+1)(\zeta_{m}^2+1)}\,.
\end{equation*}
An interesting effect happens around the point where $\tilde{\omega}_{\mathrm{e}}=\tilde{\omega}_{\mathrm{o}}$, which we dub \emph{optical coalescence} and which is characterized by the merging together of the two Lorentzian profiles [see the illustration of the mechanism in \fref{fig:CavityModes}(c)] thereby giving rise to a qualitatively different cavity response. The threshold to this behavior occurs when $\delta=\kappa$ or, in terms of the polarizabilities, when
\begin{equation*}
\zeta_\mathrm{m}=\zeta_\mathrm{m}^\ast\equiv2\zeta\sqrt{\zeta^2+1}\,.
\end{equation*}
As $\abs{\zeta_\mathrm{m}}$ is further increased beyond $\abs{\zeta_\mathrm{m}^\ast}$, the transmission attains only one peak, whose value rapidly decreases. In this regime, the cavity acts more like a single mirror whose properties are determined by that of the middle reflector than a Fabry--P\'erot cavity. A striking feature of the coalescence regime is the increased transmission bandwidth. This is reminiscent of white-light cavities familiar in the field of interferometric gravitational-wave detection~\cite{Wicht1997,*Pati2007,*Savchenkov2006}, which allow the detector to benefit from significant intracavity laser power over large bandwidths.
\par
At this point, we note that the criterion $\zeta_\mathrm{m}\sim\zeta_\mathrm{m}^\ast$ for observing coalescence-related effects is rather demanding on the quality of the middle reflector, which has to be markedly more reflective than the end-mirrors. However, this condition may be significantly alleviated through the use of a multi-layered reflector~\cite{Bhattacharya2008,Hartmann2008,Xuereb2012c}. Indeed, the effective collective polarizability of a multi-layered array scales approximately exponentially with the number of layers~\cite{Deutsch1995,Xuereb2012c}. This yields an effective coalescence threshold $\abs{\zeta_\mathrm{m}^\ast}\approx\sqrt[N]{\zeta^2/2^{N-2}}$ and
implies that each element of the middle reflector only needs to be as good as the end-mirrors for $N=2$ -- or even markedly less for $N>2$ -- in order to achieve the coalescence condition.
\begin{figure}[t]
 \includegraphics[width=0.9\columnwidth]{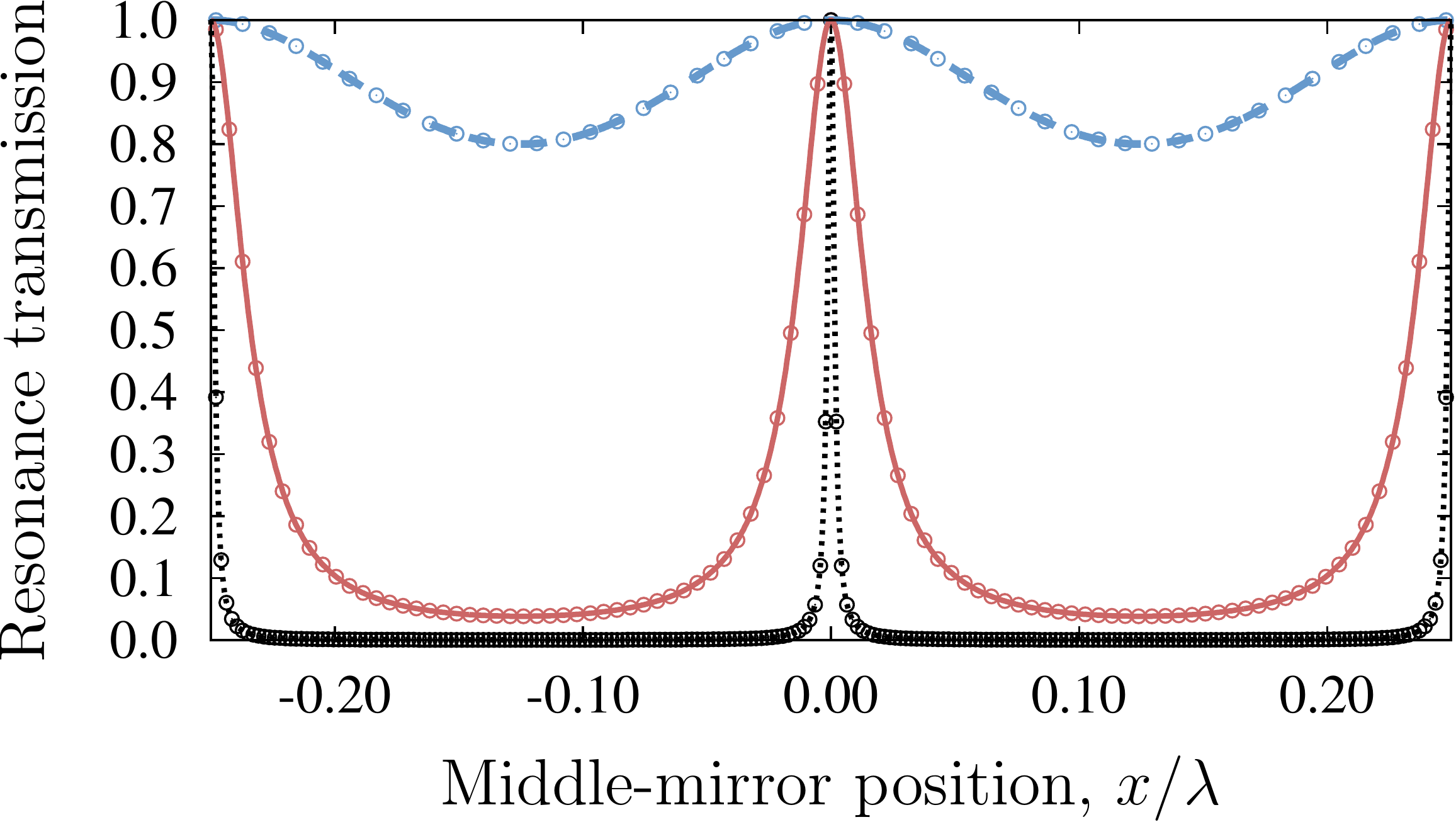}
 \caption{(Color online) The resonant transmission as a function of the membrane position, $x$. We use $\zeta=-10$ and $\zeta_\mathrm{m}=-0.5$ (blue), $-5$ (red), and $-50$ (black). The data points show the numerically-calculated transmission, and the solid curves $T_\mathrm{res}(x)$.}
 \label{fig:Transmission}
\end{figure}
\par
Let us now displace the membrane from the exact center of the cavity by $x$. One finds that, as expected, the resonance frequency depends on $x$. At each value of $x$, the resonant transmission at the respective resonant frequency, for large $\abs{\zeta}$ and below the coalescence threshold, can be approximated well by
\begin{equation}
\label{eq:Tres}
T_\mathrm{res}(x)\approx\frac{1}{1+[\zeta_\mathrm{m}\sin(4\pi x/\lambda)]^2}
\end{equation}
where $\lambda$ is the resonant wavelength, itself a function of $x$. Unity transmission can thus only be achieved when the membrane is displaced by a multiple of $\lambda/4$ and quickly drops elsewhere as $\abs{\zeta_\mathrm{m}}$ increases (\fref{fig:Transmission}).

\noindent\emph{Simple two-mode model.}---We now construct a simple two-mode Hamiltonian that provides a phenomenological basis for our discussion. Two factors are important for this discussion. First, as illustrated in \fref{fig:CavityModes}(a), adjacent resonances have opposite parities, i.e., if the field profile of one resonance is odd, that of the next is even. Second, our cavity is patently double-sided, and therefore each mode couples to two infinite baths of harmonic oscillators, one on each side of the cavity. We define the annihilation operator $b_{\mathrm{l,r}}(\omega)$ to correspond to the left (l) or right (r) bath field at a frequency $\omega$. Standard techniques and the Markovian assumption~\cite{Gardiner2004} allow us to define input (output) fields $b_{\mathrm{l,r}}^\mathrm{in(out)}(t)$ as appropriate Fourier transforms of $b_{\mathrm{l,r}}(\omega)$. A simple argument based on coordinate inversion shows that the even (odd) mode of the pair couples to an even (odd) combination of bath modes. We define two cavity modes $a_{\mathrm{o,e}}$ with corresponding frequencies $\omega_\mathrm{o,e}=\omega\pm\delta$, such that the free Hamiltonian is simply
\begin{equation*}
H_0=(\omega-\delta)a_\mathrm{o}^{\dagger}a_\mathrm{o}+(\omega+\delta)a_\mathrm{e}^{\dagger}a_\mathrm{e}\,.
\end{equation*}
We include the input noise terms for $a_{\mathrm{o,e}}$ as the two linear combinations $[b_{\mathrm{l}}^\mathrm{in}(t)\pm b_{\mathrm{r}}^\mathrm{in}(t)]\bigl/\sqrt{2}$. Assuming driving through a given mirror at a frequency $\Omega$, we can calculate the cavity transmission at the other mirror under steady-state conditions, obtaining
\begin{equation*}
T_{x=0}(\Omega)=\abs{\frac{\kappa}{\kappa+i(\omega-\delta-\Omega)}-\frac{\kappa}{\kappa+i(\omega+\delta-\Omega)}}^2\,.
\end{equation*}
This simple expression, in excellent agreement with the spectra such as those of \fref{fig:CavityModes}(c) when obtained from the full numerics, clearly shows the interference between the two Lorentzians as one approaches coalescence.

\begin{figure}[t]
 \includegraphics[width=0.9\columnwidth]{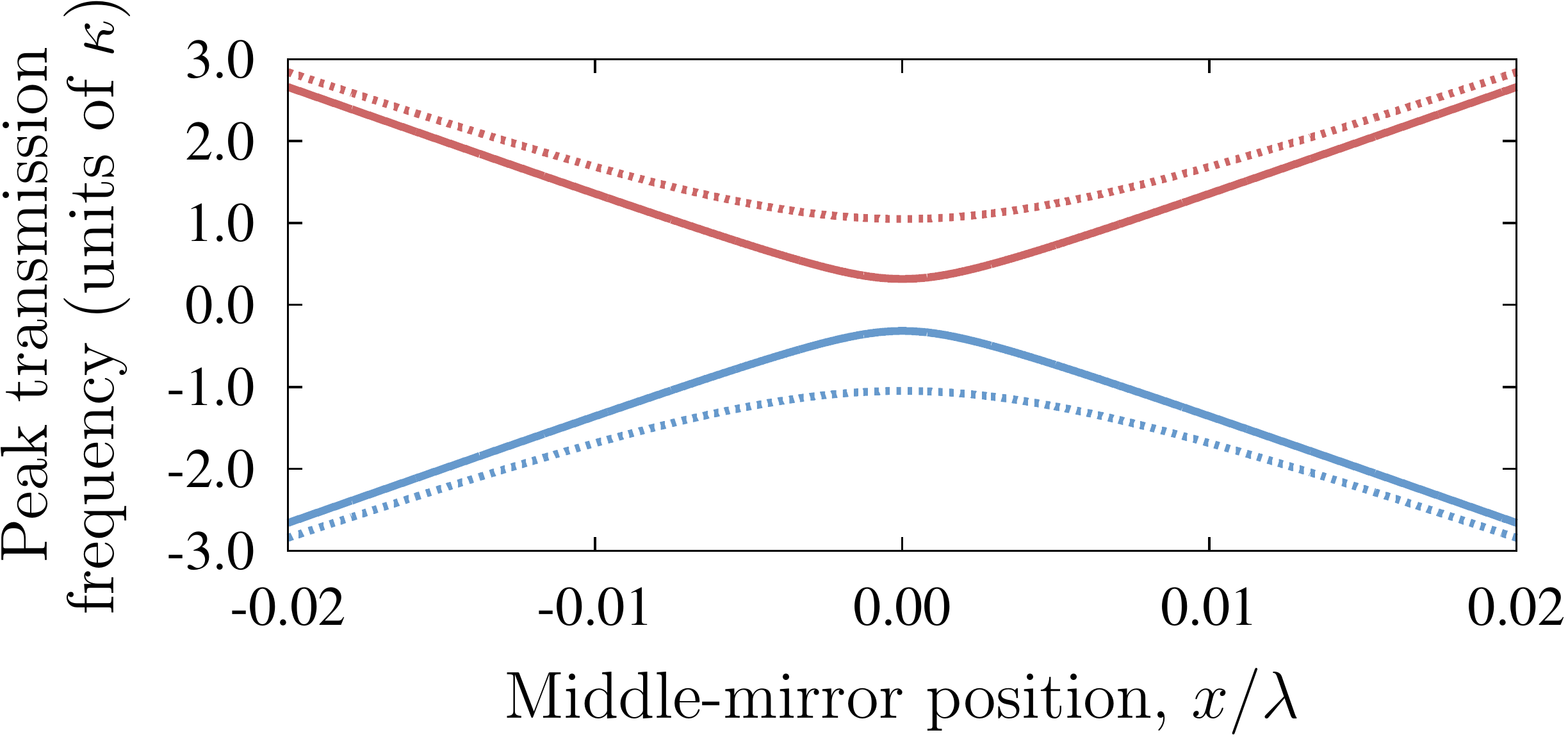}
 \caption{(Color online) Mode-pulling behavior of adjacent cavity modes ($\zeta=-10$, $\zeta_\mathrm{m}=-196.6$). Shown are two adjacent resonances as a function of middle-mirror displacement about the center of the cavity. For perfect cavity mirrors, i.e., zero linewidth, one obtains the resonances shown by the dotted curves. Imperfect cavity mirrors cause an interference effect that pulls adjacent modes closer together (solid curves), resulting in a sharper avoided crossing, and enhanced quadratic optomechanical coupling.}
 \label{fig:Quadratic}
\end{figure}

\noindent\emph{Nonlinear optomechanics.}---Let us now consider an application of optical coalescence in the field of optomechanics. When the equilibrium position of the mirror is at the centre of the cavity, the photon--phonon coupling is quadratic in $x$, corresponding to a dependence of the cavity resonance frequency on $x^2$. Such a quadratic optomechanical coupling has been proposed, e.g., to perform quantum non-demolition measurements (QND) of the phonon number operator~\cite{Braginsky1980}, possibly resolving single-phonon quantum jumps~\cite{Sankey2010,Clerk2010,Ludwig2012}. For a typical cavity operating far from coalescence, the resonance frequency can be written $\omega(x)\approx\omega+G_2^{(0)}x^2$ (where $\omega$ is the resonance at rest) and the quadratic coupling strength $G_2^{(0)}=\pm2\omega^2/(cL)\abs{\zeta_{\mathrm{m}}}$, which scales linearly with the reflector polarizability~\cite{Sankey2010,Xuereb2012d}. Membrane-optomechanics experiments~\cite{Thompson2008,Sankey2010,Karuza2013,*Purdy2013} usually operate in the regime $\abs{\zeta_\mathrm{m}}\ll\abs{\zeta}$, and, although progress in making highly reflective membranes is ongoing~\cite{Kemiktarak2012,*Kemiktarak2012b,Bui2012}, this quadratic coupling remains typically much weaker than the linear coupling.\\
In the following we exploit the coalescence mechanism to demonstrate a significantly enhanced scaling of the readout sensitivity of this quadratic coupling with $\zeta_\mathrm{m}$. To this end, we start with our two-mode model described by $H_0$ and add a displacement-induced tunneling term, governed by the tunneling strength $g_\mathrm{m}x$:
\begin{equation}
\label{eq:HamOM}
H=H_0+g_\mathrm{m}x(a_\mathrm{o}^{\dagger}a_\mathrm{e}+a_\mathrm{o}a_\mathrm{e}^{\dagger})\,.
\end{equation}
This Hamiltonian has been previously treated (for example in Refs.~\cite{Bhattacharya2008,Ludwig2012,Xu2013}) and reproduces the transfer matrix results in a quasi-static picture. The transmission peaks occur at $\omega\pm\Omega_{\pm}(x)$, with $\Omega_{\pm}=\pm\sqrt{\delta^2+(g_\mathrm{m}x)^2-\kappa^2}$, and agree with the exact numerically calculated $\tilde{\omega}_\mathrm{o(e)}$. Comparison with Ref.~\cite{Xuereb2012d} provides an expression for the tunneling parameter appearing in \eref{eq:HamOM}:
\begin{equation*}
g_\mathrm{m}=(2\omega/L)\bigl\{(\zeta_\mathrm{m}/2)\tan^{-1}\bigl[2\zeta_\mathrm{m}/(\zeta_\mathrm{m}^2-1)\bigr]\bigr\}^{1/2}\,.
\end{equation*}
One can calculate the resonant transmission of this system as a function of $x$, obtaining
\begin{equation*}
T_\mathrm{res}(x)\approx\frac{1}{1+(g_\mathrm{m}x/\delta)^2}\,,
\end{equation*}
which is consistent with \eref{eq:Tres} when expanded around $x=0$. Let us now define the sensitivity of the output field to the motion of the middle mirror as the coefficient of $x^2$ in an expansion of $\Omega_\pm(x)$ around $x=0$:
\begin{equation*}
\Omega_\pm(x)=\Omega_\pm(0)+G_2(\zeta_\mathrm{m})x^2+\dots\,.
\end{equation*}
In the limit of small displacements, $g_\mathrm{m}\abs{x}\ll\sqrt{\delta^2-\kappa^2}$, one finds that
\begin{equation*}
G_2(\zeta_\mathrm{m})=\frac{2\zeta^2}{\sqrt{\zeta^{*2}_\mathrm{m}-\zeta_\mathrm{m}^2}}G_2^{(0)}\,.
\end{equation*}
In the two-fold limit of good cavity mirrors ($\abs{\zeta}\gg1$) and away from coalescence ($\abs{\zeta_{\mathrm{m}}}\ll\zeta^2$), this expression reduces to the coupling strength: $G_2(\zeta_\mathrm{m})\to G_2^{(0)}$. This corresponds to the intuitive statement, which we emphasize is correct \emph{only away from the coalescence condition}, that the readout sensitivity of the system is given by the optomechanical coupling strength.\\
As one approaches coalescence, however, one observes a dramatic increase in the readout sensitivity. However, whilst the readout sensitivity is enhanced by a factor $G_2(\zeta_\mathrm{m})/G_2^{(0)}$, the backaction of the field on the middle mirror is still quantified through the interaction strength $G_2^{(0)}$ and, therefore, \emph{unmodified}. The physical mechanism at the basis of this enhancement is the coalescence-induced mode-pulling mechanism illustrated in \fref{fig:Quadratic}, where the presence of a nearby mode `pulls' a second mode towards the former. However, as can be seen from the expression for $\Omega_{\pm}(x)$, the displacement region over which this enhancement is obtained becomes smaller, and ultimately vanishes, as one approaches coalescence. One can easily show that the quadratic expansion is valid as long as $G_2(\zeta_\mathrm{m})/G_2^{(0)}\ll 2/(\eta\abs{\zeta_\mathrm{m}})$, where $\eta=2\pi\abs{x}/\lambda$ is the Lamb--Dicke parameter. For typical membrane-based optomechanical experiments experiencing zero-point fluctuations over $\sim$\,fm length-scales and for $\sim\!1\,\upmu$m wavelengths, the enhancement can still be of several orders of magnitude. For a two-membrane stack, each having mass $\sim\!100$\,ng, frequency $2\pi\times 100$\,kHz and $\zeta_\mathrm{m}\approx-10$, the enhancement reaches $\sim\!10^6$ for ground state-cooled membranes and $\sim\!10^3$ for membranes at $4$\,K.\par
We stress that the enhancement found in this work is a consequence of the interference in the output fields and arises purely from a full treatment of the \emph{static} properties of the system. This is in contrast with the dynamical effects discussed in Refs.~\cite{Ludwig2012,Stannigel2012,Xu2013}, in which the application of the Hamiltonian in \eref{eq:HamOM} to typical optomechanical situations with mechanical resonators in the good-cavity limit can be used to achieve enhanced nonlinearities and displacement sensitivities. We nevertheless note that a combination of such dynamical effects with the mechanism reported here could give rise to interesting behavior in the regime around coalescence, in both the good- and bad-cavity limits.

\noindent\emph{Concluding remarks.}---We have explored a fundamental effect in the electrodynamic description of a cavity with
a partially-transmitting mirror inside it, in the regime where this mirror is significantly better than the cavity mirrors themselves. Under these conditions, interference between the output fields for adjacent cavity modes causes a significant enhancement to the readout sensitivity of the motional coordinate of the middle mirror, without an associated enhancement to the backaction of the field on this coordinate.

We acknowledge support from the Austrian Science Fund (FWF):\ P24968-N27 and an STSM grant from the COST Action MP1006 ``Fundamental problems in quantum physics'' (C.G.), the Royal Commission for the Exhibition of 1851 (A.X.), the Universit\'{e} de Strasbourg through Labex NIE, IDEX, and EOARD (G.P.), and the Danish Council for Independent Research under the Sapere Aude program (A.D.).

\end{document}